\documentstyle[amssymb,prl,aps,epsf]{revtex}
\begin{document}
\draft

\twocolumn[\hsize\textwidth\columnwidth\hsize\csname
@twocolumnfalse\endcsname
\title{Vortex Lattice Melting into Disentangled Liquid Followed by the 3D-2D
Decoupling Transition in YBa$_2$Cu$_4$O$_8$ Single Crystals }
\author{X.G. Qiu\cite{byline}, V.V. Moshchalkov, and Y. Bruynseraede}
\address{Laboratorium voor Vaste-Stoffysica en Magnetisme, Katholieke Universiteit Leuven, \\
Celestijnenlaan 200D, B-3001 Leuven, Belgium}
\author{J. Karpinski}
\address{Laboratorium f\"{u}r Festk\"{o}rperphysik, ETH-H\"{o}nggerberg, CH-093,
Z\"{u}rich, Switzerland}
\maketitle

\begin{abstract}
A sharp resistance drop associated with vortex lattice melting was observed
in high quality YBa$_2$Cu$_4$O$_8$ single crystals. The melting line is well
described well by the anisotropic GL theory. Two thermally activated flux
flow regions, which were separated by a crossover line B$_{cr}$(T)=1406.5(1-T/T$_c$)/T (T$_c$=79.0 K, B$_{cr}$ in T), were observed in the
vortex liquid phase. Activation energy for each region was obtained and the
corresponding dissipation mechanism was discussed. Our results suggest that
the vortex lattice in YBa$_2$Cu$_4$O$_8$ single crystal melts into
disentangled liquid, which then undergoes a 3D-2D decoupling transition.
\end{abstract}
\vspace{-20pt}
\vskip2pc] \vskip2pc \narrowtext

The vortex dynamics in the mixed state of high T$_c$ superconductors (HTSCs)
remains a subject of intense research because of its fundamental importance
for physics of the vortex matter\cite{ref1,ref2,ref3}. It was predicted
theoretically\cite{ref4,ref5,ref6} and confirmed experimentally\cite
{ref7,ref8,ref9,ref10,ref11,ref12,ref13,ref14,ref15} that, the vortex system
undergoes a second order transition from a high temperature vortex liquid to
a low temperature vortex glass or Bose glass in superconductors with strong
disorders. In a clean superconductor a first order transition from vortex
liquid to a well ordered vortex lattice occurs.

Vortex melting in high quality high Tc superconducting single crystals has
been detected by various techniques including torsion oscillator\cite{ref8},
transport\cite{ref9,ref10}, magnetization\cite{ref11,ref12}, neutron
scattering\cite{ref13}, $\mu $ spin rotation\cite{ref14} and calorimetric
measurements\cite{ref15}. Central to the question of flux melting is: does
the vortex lattice melt via a first order transition? How does disorder
influence the melting transition\cite{ref16}? It was found by Safar et al.
\cite{ref9} and by Kwok et al.\cite{ref10} that the resistive transition of
high quality YBa$_2$Cu$_3$O$_7$ (Y123) single crystal showed a sharp
discontinuity and that the transition exhibited a hysteric behavior. They
attributed this phenomenon to flux melting. Although their explanation was
questioned by Jiang et al.\cite{ref17} who found that the hysteresis was not
necessarily an indication for a melting transition, it could be resulted
from a nonequilibrium behavior seen in the resistive transition. Subsequent
magnetization\cite{ref11,ref12} and calorimetric\cite{ref15} measurements
convincingly proved that the melting transition was indeed a first order
one, i.e., discontinuity in the internal energy at the melting temperature
was observed. By doing simultaneous transport and magnetization
measurements, Welp et al.\cite{ref12} and Fuchs et al.\cite{ref18} found
that the temperatures, where the resistance jumped, coincided with the
melting temperature obtained from magnetization measurements on Y123 and Bi$%
_2$Si$_2$CaCu$_2$O$_8$ (Bi2212) single crystals, respectively. The role of
disorder in the melting transition has been checked by Kwok et al.\cite
{ref10} and by Fendrich et al.\cite{ref19}. They found that strong pinning
from twin boundaries or artificially introduced defects drove the first
order melting transition into a second order glass transition.

Another important question related to the flux melting is how the melting
propagates. Will the vortices lose their coherence along the c direction
during the melting transition or will they keep the c-axis integrity and
then lose it afterwards, at higher temperature, when thermal fluctuations
destroy long range correlations along the c-direction\cite{ref20}? This
question has been recently pursued by employing the flux transformer
configuration. The results remain still controversial. Lopez et al. and
Fuchs et al. found a coincidence of the melting transition and decoupling
one in Y123 and Bi2212 single crystals, respectively\cite{ref21}. However,
Wan et al. and Keener et al. did similar measurements on Bi2212 single
crystals and concluded that the melting took place in a two stage fashion
\cite{ref22}.

Previous studies of the vortex melting transition have been concentrated on
Y123 and Bi2212 single crystals. Up to now, no detailed measurements on YBa$%
_2$Cu$_4$O$_8$ single crystals have been reported. Meanwhile, the
dissipation in the liquid state after the melting transition still remains
to be clarified. The main purpose of this work is to show that in YBa$_2$Cu$%
_4$O$_8$ single crystals, vortex lattice melts into liquid via a first order
transition. It was also found that the liquid state after the melting
transition could be separated into two regions which were governed by
different dissipation mechanisms. Our results support the picture that {\it %
the vortices first lose their transverse coherence at the melting
temperature, this lose being followed by a disappearance of correlation in
longitudinal direction at higher temperature}.

YBa$_2$Cu$_4$O$_8$ (Y124) single crystals were grown by travelling solvent
floating zone method. The details for the single crystal growth were
published elsewhere\cite{ref23}. These single crystals are needle-like with
typical dimensions of 1.2$\times $0.3$\times $0.05 mm$^3$. T$_c$ of the
single crystals was about 76-78 K. Three single crystals were used for
similar measurements. Each crystal was carefully cleaved to obtain optically
flat surfaces with the c-axis normal to the sample surface. Gold wires were
attached to the surface of crystal by using Pt epoxy. Then the crystal was
heated in air at 100 $^{\circ }$C for 1 hour, yielding a typical contact
resistance below 0.5 $\Omega $. The resistance was measured by four-probe
low frequency ac lock-in technique with an excitation current of 0.3 mA at
17 Hz in the ab plane. The magnetic field was generated by a 15 T Oxford
superconducting magnet. The direction between the crystalline c-axis and the
magnetic field was adjusted by rotating the sample holder with an accuracy
of 0.2 $^{\circ }$. The crystal we report on here had a T$_c$ of 77.6 K and
transition width of about 1.2 K at zero field.

\begin{figure}[tbp]
\centerline{
\epsfxsize=3.0 in
\epsfbox{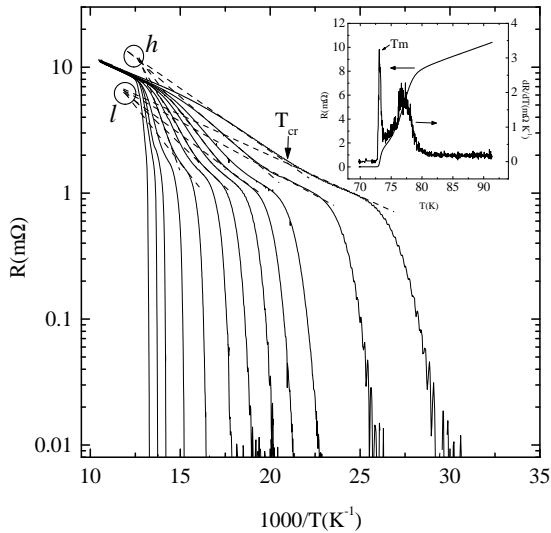}}
\caption{Arrenhius plot of resistance vs. 1000/T at different applied
magnetic fields. Solid lines are experimental data. From left to right: H=
0.2, 0.5, 1, 2, 3, 4, 5, 6, 7, 8, 10, 12 T. The dashed lines are the fits
with TAFF theory. $h$ and $l$ are used to identify the two TAFF regions. T$%
_{cr}$ shows how the decoupling transition temperature is determined. Inset:
determination of T$_m$ from the dR/dT vs. T plot derived from the R-T curve
for H=0.5 T.}
\label{fig1}
\end{figure}

Fig.1 shows the Arrenhius plot for the temperature dependence
of the resistance measured at different fields applied perpendicular to the
ab plane up to 12 T. A sharp jump in the resistance R with a magnitude of R/R%
$_{n}\sim $10\% is clearly visible at each magnetic field. The magnitude of
the resistive drop lies between those of Y123 ($\thicksim $20\%) and Bi2212 (%
$\sim $0.5\%), suggesting a possible dependence of the R(T) jump on the
anisotropy of the materials since Y124 has an anisotropy between those of
Y123 and Bi2212. We also observed small hysteresis in the resistive jump
upon warming up and cooling down. We measured resistive transition at 4 T
with different excitation currents of 0.1 mA, 0.3 mA and 1 mA, respectively.
The R-T curves measured at 0.1 mA and 0.3 mA were essentially identical.
However, that of 1 mA deviated from those of 0.1 mA and 0.3 mA. Similar
hysteric resistance jumps were earlier observed on Y123 and Bi2212 single
crystals and were attributed to a first order flux melting in clean
superconductors\cite{ref9,ref10}. To quantify the melting transition, we
determine the melting temperature T$_{m}$ as the temperature where dR/dT vs.
T plot shows a sharp peak. The inset in Fig. 1 demonstrates how the melting
points T$_{m}$ are determined. The obtained melting temperatures are plotted
against the corresponding magnetic fields in Fig. 2. The melting line can be
described by the empirical formula B$_{m}$(T)$\sim $(1-T/T$_{c}$)$%
^{\alpha }$ with the exponent $\alpha $ between 1 and 2\cite{ref8,ref9,ref10}%
. We find that the melting line is best fitted by B$_{m}$(T)=31.51(1-T/T$_{c}
$)$^{1.47}$, T$_{c}$=77.6 K being the transition temperature at zero field.
This result agrees well with that obtained by Kwok et al.\cite{ref10} who
found B$_{m}$(T)=103(1-T/92.33)$^{1.41}$ on Y123 single crystal when the
field was applied parallel to c-axis.

\begin{figure}
\centerline{
\epsfxsize=3.0 in
\epsfbox{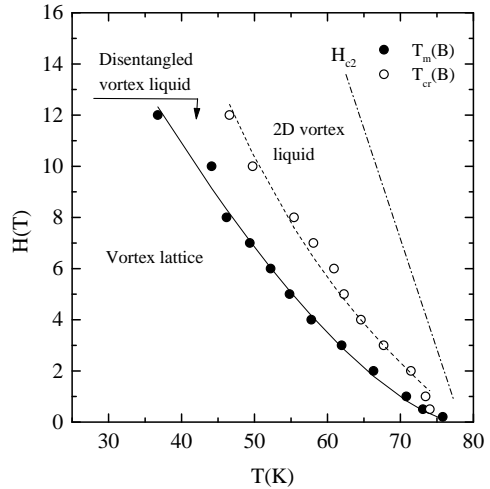}}
\caption{The H-T phase diagram for vortices in YBa$_2$Cu$_4 $O$_8$ single
crystal. Solid line and dashed line are best fittings to the experimentally
obtained melting line (filled circles) and 3D-2D decoupling line (open
circles) with B$_m$(T)=31.5(1-T/77.6)$^{1.47}$ (solid line) and B$_{cr}$%
(T)=1406.5(1-T/79.0)/T (dashed line), respectively.}
\label{fig2}
\end{figure}

To check further an interpretation of the B$_m$(T) line in the framework of
the vortex melting scenario, we measured the resistive transition with the
magnetic fields applied at different angles to the ab plane, while fixing
the field magnitude at 4 T. The angular dependence of the melting transition
is shown in Fig.3 where clear resistance jumps can be seen at all the
angles. It is interesting to note that at small angles ($\theta $%
\mbox{$<$}%
20$^{\circ }$, $\theta $ being the angle between ab plane and applied
field), the resistive transition shows an anomalous behavior above the kink.
This anomaly was more clearly visible in our flux flow measurements (R-B
curves). Such kind of anomalous behavior could be due to the vortex tilting
instability at large angles and under a magnetic field of a few Tesla which
induces a competition between intrinsic pinning and vortex interaction.
Following the same procedure, we have obtained the angular dependence of the
melting transition temperatures at 4 T as shown in the inset of Fig. 3.

For a 3D vortex lattice, the melting transition occurs when the shear
modulus $c_{66}$ goes to zero, which is determined by using a
phenomenological Lindermann criterium, i.e., when the mean square root
amplitude $\sqrt{\langle u^2\rangle }$ of the displacement of the vortex
from its equilibrium position is larger than a certain portion of the
lattice constant, $\sqrt{\langle u^2\rangle }>c_La_0$ where $c_L$ is the
Lindermann number, $a_0$ being the vortex spacing. According to the theory
for an anisotropic superconductor, T$_m$($\theta $) has such a relationship
as\cite{ref6} 

\begin{figure}
\centerline{
\epsfxsize=3.0 in
\epsfbox{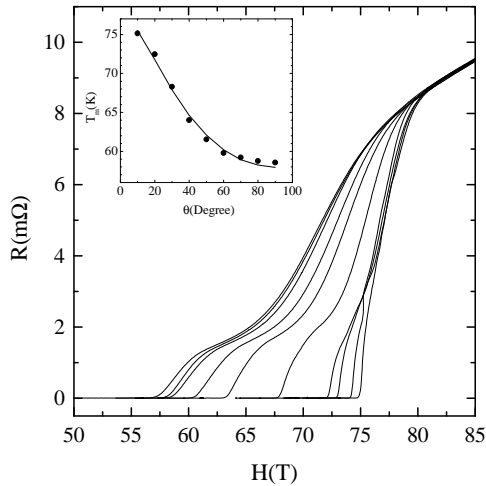}}
\caption{Angular dependence of the resistive transition at a fixed magnetic
field of 4 T. From left to right: $\protect\theta $=90, 70, 60, 50, 40, 30,
20, 10, 5, 2 $^{\circ}$. Inset: Angular dependence of the melting transition
in YBa$_2$Cu$_4$O$_8$, solid line is fitting by using anisotropic GL theory
Eq. (1) with a Lindermann number $c_L$=0.14.}
\label{fig3}
\end{figure}

\begin{equation}
k_BT_m=\frac{\Phi _0^{5/2}c_L^2}{4\pi ^2\lambda
_{ab}^2(T_m)B^{1/2}\varepsilon ^{1/4}(\theta )}\text{ ,}  \label{eq1}
\end{equation}
where $\varepsilon (\theta )=\cos ^2(\theta )+\gamma ^2\sin ^2(\theta )$, $%
\Phi _0=2.07\times 10^{-7}G$ $cm^2$ is the flux quantum, $\lambda _{ab}$ is
the penetration depth in the ab plane and $\gamma =\lambda _c/\lambda _{ab}$
the anisotropy parameter. A best fit to the obtained data by using Eq. (\ref
{eq1}) with $\lambda _{ab}=2000$ \AA\ (Ref. 24) is achieved when $\gamma $%
=13.6, $c_L=0.14$, which is shown as the solid line in the inset of Fig.3.
Clearly a quite satisfactory result has thus been obtained. The $c_L$ value
agrees very well with those reported by Safar et al.\cite{ref9} and by Kwok
et al.\cite{ref10}. Considering the crystalline structures of Y123 ($\gamma $%
=7.7)\cite{ref10}, Y124 and Bi2212 ($\gamma $=50$\sim $170)\cite{ref25}%
, we think that the anisotropy of $\gamma $=13.6 is a reasonable value.

As mentioned in the introductory part, in a clean layered superconductor,
the melting line separates the vortex lattice phase at low temperatures from
the vortex liquid phase above B$_m$(T). Depending on the temperature,
magnetic field and strength of disorder, the vortex can be melted into a
disentangled 3D line liquid or an entangled liquid\cite{ref1,ref26}. It is
important to note that the melting transition is not necessarily a depinning
transition, which means that vortex liquid can be pinned. The pinning force
can arise from the residual pinning or from viscosity of the vortices. Due
to the different dimensionality and vortex configurations involved in these
various liquid states, the dissipation mechanisms should also be quite
different. A useful information about the liquid state can be extracted by
carefully analyzing the activation behavior of vortex liquid.

\begin{figure}
\centerline{
\epsfxsize=3.0 in
\epsfbox{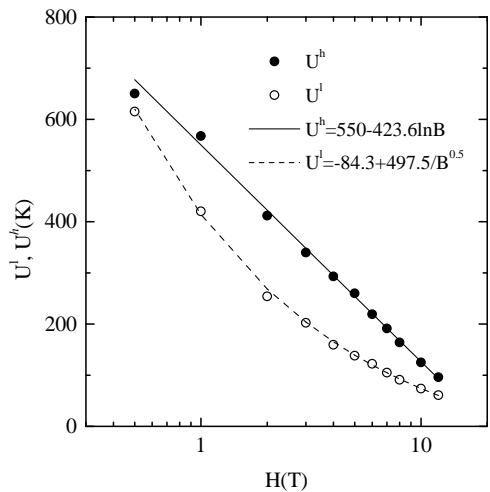}}
\caption{Activation energies derived from the slopes in Fig.1. Filled circle
and open circles are experimental data. Solid line and dashed line are the
corresponding best fits.}
\label{fig4}
\end{figure}

From Fig. 1 it is clearly seen that above the melting temperature, i.e. in
the liquid state, there are two distinct parts where lnR shows a linear
dependence upon 1/T with different slopes. This kind of behavior ($\rho
=\rho _0e^{-U/kT}$) is typical for thermally activated flux flow (TAFF) with
an activation energy $U(B,T)=U(B,0)(1-T/T_c)$\cite{ref27}. Such two separate
TAFF regions with different activation energies have previously been
observed by us on oxygen deficient Y123 thin films\cite{ref27}.

Using the same notations as in Ref. 27, we define the TAFF regions as region 
$l$ and region $h$ as shown in Fig.1. For each TAFF region, we extract the
activation energies from the slopes of the Arrenhius plot. The obtained
results are shown in Fig. 4. We find that the activation energy can be best
fitted by $U_0^l=-84.3+497.5/\sqrt{B}$ and $U_0^h=550-423.6\ln B$,
respectively. Moreover, from a low temperature extrapolation of TAFF fit to
region $h$ and a high temperature one for region $l$, we obtain the
temperatures $T_{cr}(B)$ which define the crossover from region $l$ to $h$.

It was argued\cite{ref1,ref26} that the Abrikosov lattice could melt into a
disentangled liquid under certain circumstances. As discussed by Geshkenbein
et al. and by Vinokur et al., the dissipation in a disentangled liquid can
be developed via plastic deformations of the vortices\cite{ref28}. The
activation energy is associated with the energy required to create a double
kink in the vortex which is the free energy of two vortex segments along the
ab plane with the length of $a_0=(\Phi _0/B)^{1/2}$, the average distance
between the vortices. Using anisotropic Ginzburg-Landau theory, the barrier
for this plastic movement can be estimated as 
\begin{equation}
U_{pl}=2E_va_0\simeq \frac{\Phi _0^2}{8\pi ^2\gamma \widetilde{\lambda }^2}%
\left( \frac{\Phi _0}B\right) ^{1/2}\text{ ,}  \label{eq2}
\end{equation}
where $E_v$ is the vortex energy per unit length along the ab planes and $%
\widetilde{\lambda }^2=\lambda _{ab}\lambda _c$. With $\lambda
_{ab}^2=\lambda _{ab}^2(0)/(1-t)$(where $t=T/T_c$), an activation energy $%
U_0=U_{pl}\propto (1-t)/\sqrt{B}$ is predicted which is in qualitative
agreement with our experimental data. To make an order of magnitude
estimation of $U_{pl}$, we have used the above value for $\gamma $(=13.6)
and $\lambda _{ab}=2000$ \AA , thus obtaining $U_0\approx 1500$ $K$ at 1 T,
which agrees well with the experimental data.

As the temperature increases, due to the enhanced thermal fluctuations, an
entanglement-disentanglement or 3D-2D decoupling transition should happen.
Considering the high temperature, we think a 3D-2D decoupling is more
likely. Daemen et al.\cite{ref20} have calculated self-consistently the
decoupling line for a Josephson coupled superconducting system. When the
renormalization of the Josephson coupling by thermal fluctuations and static
disorder are taken into account, the decoupling field is given by 
\begin{equation}
B_{cr}(T)=\frac{\Phi _0^3}{16\pi ^3k_BTse\lambda _{ab}^2(T)\gamma ^2}\text{ ,%
}  \label{eq3}
\end{equation}
for moderate anisotropy when $\xi _{ab}\ll \gamma s\ll \lambda _{ab}$, where
s is the distance between the superconducting Cu-O planes ($s\sim 10$ \AA\
for Y124)\cite{ref24}. Since $\lambda _{ab}^2=\lambda _{ab}^2(0)/(1-t)$, we
have $B_{cr}(T)\propto (1-T/T_c)/T$. We find that the $B_{cr}(T)$ line we
obtained above can be nicely fitted by Eq. (\ref{eq3}). The fitted result is 
$B_{cr}=1406.5(1-T/T_c)/T$ as shown in Fig. 2 by the dashed line.

As discussed in Ref. 27, the dissipation in the 2D liquid state is
controlled by the plastic motion of pancake vortices which can be visualized
as the generation of dislocation pairs. The typical energy for creating such
a pair is 
\begin{equation}
U_e=\frac{\Phi _0^2s}{16\pi ^3\lambda _{ab}^2(T)}\ln (B_0/B)  \label{eq4}
\end{equation}
with $B_0\simeq \Phi _0/\xi ^2(T)$. We note that Eq. (\ref{eq4}) gives a
(1-t)lnB dependence of $U_e$ as we have just observed. Inserting the values
for s ($\sim $10 $\AA $ ) and $\lambda _{ab}$($\sim $2000 $\AA $ ), we find $%
U_e=146.5(lnB_0-lnB)$, in excellent agreement with our experimental data.
The large prefactor $\rho _0$ derived from TAFF analysis for region $h$
suggests that the liquid in region $h$ is very viscous. Since a 2D pancake
vortex liquid is just an extreme case for an entangled liquid, i.e., the
entanglement length equals to $s$. The entanglement increases the viscosity
of vortex liquid and results in a larger activation energy, and thus a large
prefactor.

Therefore, our results suggest feasibility of the following scenario: below $%
B_m(T)$, the vortices form a regular Abrikosov lattice. {\it As the
temperature increases, the lattice melts into a disentangled liquid, loses
its transverse correlation while keeping the longitudinal one. The
dissipation is governed by the generation of double kinks. Upon further
warming, a 3D-2D decoupling transition sets in and the vortices lose their
longitudinal coherence. The 2D vortices form very viscous liquid and the
activation energy increases which follows a }$(1-t)lnB${\it \ relationship}.

In conclusion, we have shown that vortex lattice melting can also been
observed in Y124 single crystals. The melting line can be well described by
using anisotropic GL theory with a Lindermann constant $c_L=0.14$. Two
different liquid phases were observed which we think correspond to the
disentangled liquid and the 2D pancake vortex liquid, respectively.

We thank E. Rossel and P. Wagner for their help during the measurements.
This research has been supported by the Belgian IUAP and Flemish GOA and FWO
programs.

\end{document}